\documentclass[twocolumn,showpacs,prl]{revtex4}
\usepackage{amsmath,amssymb}
\usepackage{amsfonts}

\def \w {\omega} 
\def \I {\mathfrak {I}}

\def \M {\mathcal {M}} 

{
\begin{document}
\title{Quantization of horizon areas of the Kerr black hole}

\author{Yongjoon Kwon }
\email{emwave@khu.ac.kr}

\author{Soonkeon Nam }
\email{nam@khu.ac.kr}

\affiliation{Department of Physics and Research Institute of Basic Sciences, Kyung Hee University, Seoul 130-701, Korea}

\date{\today}

\begin{abstract}
Since the Bekenstein's proposal that a black hole has equally spaced area spectrum, the quasinormal modes as the characteristic modes of a black hole have been used  in obtaining the horizon area spectrum of the black hole. However, the area spectrum of the Kerr black hole in some previous works was inconsistent with the Bekenstein's proposal.
In this paper, we notice that the Kerr black hole has three types of resonance modes which are quasinormal modes (QNM), total transmission modes (TTM), and total reflection modes (TRM). All of these resonances represent highly damped oscillations and only depend on the black hole parameters which are mass, charge and angular momentum of the black hole. We propose that all of these modes should be used in quantizing the black hole. 
With all these modes, we find that both the inner and outer horizon areas of the Kerr black hole are equally spaced.
Similar behavior is also found in the Reissner-Nordstr{$\rm {\ddot {o}}$}m black hole.
\end{abstract}
\maketitle

\textit{Introduction}---
As a quantum property of black holes, it is believed that the black hole horizon area is quantized. 
It was first proposed that the horizon area is an adiabatic invariant and should be quantized, based on Ehrenfest principle  that any classical adiabatic invariant corresponds to a quantum entity with discrete spectrum  \cite{be}. By considering the minimum change of the horizon area in the process of the assimilation of a test particle into a black hole, it was obtained that the area spectrum should be linearly quantized, i.e. $A= \gamma n \hbar$ where $\gamma$ is an undetermined dimensionless constant  \cite{be}.
When a perturbation on a black hole is given, the black hole undergoes damped oscillations which are called  quasinormal modes. 
By using these quasinormal modes of a black hole, it was realized that the area spectrum of the black hole can be obtained \cite{hod, kun, setvag, magi, vage, med, wei}.
By considering the real part of the asymptotic quasinormal modes of a black hole as a transition frequency in the semiclassical limit, the area spectrum of the Schwarzschild black hole was obtained as $A = (4\, {{\rm ln} 3}) \, n \hbar$  \cite{hod, kun}.  
For a rotating black hole an adiabatic invariant $\mathcal I $ was considered and quantized via the Bohr-Sommerfeld quantization; $\mathcal I= \int {{dM -\Omega \, dJ}  \over {\w_R} }=n \hbar$ \cite{setvag}.
As an example,  the Kerr black hole was considered and  the real part of the asymptotic quasinormal modes  given by \cite{hodk}
\begin{equation} \label{kqn}
\w  = m \Omega - i  2 \pi T_H k 
\end{equation}
 was used, where $m$ is the azimuthal harmonic indices, $\Omega$ is the angular velocity at horizon, and $k$ is overtone number with $k \in N$ and $k \gg 1$.
However it was obtained that the area spectrum was not equally spaced, which is in contradiction to the Bekenstein's proposal \cite{be}. 
Later, it was proposed that a perturbed black hole should be described as a damped harmonic oscillator and
a transition in a black hole should be considered as the transition between quantum levels $(\w_{0})_k\equiv ( \sqrt{\w_{R}^2 +\w_I^2 } )_k$, where $\w_R$ and $\w_I$ are the real and imaginary parts of the asymptotic quasinormal modes \cite{magi}. Therefore the characteristic classical frequency $\w_c$  should be identified with the transition frequency between the quantum levels $(\w_{0})_k$  in the semiclassical limit, i.e. $\w_{c} = (\w_ {0})_{k} - (\w_ {0})_{k-1} = ( | \w_ {I} | )_{k} - ( | \w_ {I} | )_{k-1} $ for the highly damped quasinormal modes, where $k \in N , k \gg 1$  \cite{magi}.
By using this transition frequency $\w_c$, the area spectrum of the Schwarzschild black hole was obtained  as $A=8 \pi n \hbar $ \cite{magi, med, wei}.
For the rotating black hole this transition frequency $\w_c$ instead of $\w_R$ was applied to the adiabatic invariant $\mathcal I$ which was introduced in Ref.\cite{setvag}.
In the works of Refs.\cite{med, vage} the transition frequency $\w_c$ of the Kerr black hole was obtained from the different calculation of the asymptotic quasinormal modes in Refs.\cite{keshk, keshk2}, which are given by 
\begin{equation} \label{keqnm2}
\w \simeq   \tilde \w _0 - i  2 \pi T_H ^s \left(k+{1 \over 2} \right) ~,
\end{equation}
where $\tilde \w _0 $ is a function of the black hole parameters whose real part asymptotically approaches ${\rm Re}(\tilde \w_0) \propto m$ and $T_H ^s = {1 /({8 \pi M} )}$. 
These quasinormal modes are different from the previous one (\ref{kqn}).
From the quasinormal modes (\ref{keqnm2}), the transition frequency $\w_c=2 \pi T_H ^s$ was obtained in the semiclassical limit, and used for Bohr-Sommerfeld quantization of the adiabatic invariant, $ I= \int {{dM -\Omega \, dJ}  \over {\w_c} }=n \hbar$. 
But this quantization condition does not give  equally spaced area spectrum for the Kerr black hole \cite{vage, med},  still  inconsistent with the Bekenstein's proposal. Only for slowly rotating case with small angular momentum $J$ compared to  mass  $M$  of the Kerr black hole, it was obtained that the area spectrum was approximately equally spaced as $A= 8 \pi n \hbar$ \cite{med, myung}.
In our recent work \cite{sk} it was reminded that an action variable is adiabatic invariant, but not every adiabatic invariant is an action variable, and that only action variable can be quantized via the Bohr-Sommerfeld quantization in the semiclassical limit.  Therefore not an adiabatic invariant but an  action variable of the classical system  should be identified in order to  apply  the Bohr-Sommerfeld quantization  \cite{sk}. 
By Bohr's correspondence principle which says that the transition frequency at large quantum number equals to the classical oscillation frequency of the corresponding classical system, the black hole with the transition frequency $\w_c$ can be considered  as  the  classical system of periodic motion with oscillation frequency $\w_c$ in the semiclassical limit. Therefore the action variable of the classical periodic system with the oscillation frequency $\w_c$ was identified and finally quantized via the  Bohr-Sommerfeld quantization in the semiclassical limit as follows \cite{sk}:
\begin{eqnarray} \label{myf}
{\I} =\int { dE \over {\w _c }} =\int { dM \over {\w _c }} = n \hbar ~~, ~~(n \in Z, \vert n \vert \gg 1)~,
\end{eqnarray}
where   the transition frequency $\w_c$ in the semiclassical limit is given by $\w_{c} = ( | \w_ {I} | )_{k} - ( | \w_ {I} | )_{k-1} $ for highly damped modes, and the change of the energy $E$ of a black hole is considered as the change of the ADM mass $M$ (or ADT mass $\M$ according to the gravity theory \cite{sk2}).
This formula can be also applied for a rotating  black hole with a transition frequency. 
For example, BTZ black hole and  warped ${\rm AdS}_3$ black hole were considered and their area and entropy spectra were obtained \cite{sk, sk2}.
From the results, it was noticed that the entropy spectrum of a black hole is more fundamental than the area spectrum, and there is the universality that the entropy spectrum of a black hole is equally spaced \cite{sk2}.

In this paper, we would like to apply the formula (\ref{myf}) for the Kerr black hole and to find if the entropy spectrum has the universal behavior of the equally spaced spectrum.  
Until now, in spite of the several attempts \cite{setvag, med, vage}  for the area spectrum of the Kerr black hole, the results  were not consistent with Bekenstein's original proposal. 
Furthermore,  for the Kerr black hole  there are two different calculations (\ref{kqn}) and (\ref{keqnm2}) of the quasinormal modes which give different transition frequencies. Therefore it is not clear which one is right for the quantization of the Kerr black hole.
Recently, by considering the scattering problem on the Kerr black hole the highly damped quasinormal modes (QNM) of the Kerr black hole were obtained from the poles of the transmission and reflection amplitudes \cite{keshk2}. However we notice that there are other highly damped oscillation modes of the Kerr black hole, which are total transmission modes (TTM) and total reflection modes (TRM) obtained in Ref.\cite{keshk2}. The TRM and TTM are obtained from the zeros of the transmission and reflection amplitudes, respectively. 
In order to find out the property of a black hole we have to do scattering experiment on the black hole.
When we consider  scattering problem on a black hole, the black hole can have other resonance modes as well as QNM.
In scattering problem of a black hole, the wave equation becomes Schr${\rm \ddot o}$dinger-like equation in quantum mechanics. Then the incident waves from infinity are reflected and transmitted because of the effective potential which plays a role of potential barrier in quantum mechanics. 
The QNM only depend on the black hole parameters like mass, charge, and angular momentum  which characterize a black hole.  
Therefore it has been considered that QNM are the characteristic modes of the black hole as a fingerprint in directly identifying the existence of a black hole \cite{nolkz} and  carry some information about quantum structure of  the black hole \cite{magi}. 
%
We notice that TTM and TRM  are also the characteristic modes of a black hole, since they  are also damped oscillations which only depend on black hole parameters. 
In this sense we should also promote  TTM and TRM to the equivalent position of QNM, so that they play the same role as QNM in quantizing the black hole.
Therefore  we will obtain the transition frequencies corresponding to each of all of them in the semiclassical limit.
%
By applying the formula (\ref{myf}) with these transition frequencies for the Kerr black hole, we  find that the area spectrum  is consistent with the Bekenstein's proposal \cite{be} and the entropy spectrum has the universality  of equally spaced spectrum.
Before that, we would like to reconsider the case of the Schwarzschild black hole.
For that case, in the highly damped regime we can find QNM and TRM, but there is no TTM. The QNM and TRM are easily obtained from the transmission and reflection amplitudes obtained in \cite{neitz} as follows:
\begin{eqnarray} 
\w^{QNM} &=&  T_H ^s  {\rm ln} 3 +i 2 \pi  T_H ^s  \left(k+{1 \over 2} \right) ,  \\
\w^{TRM} &=& i 2 \pi  T_H ^s  k  ,
\end{eqnarray}
where $ k \gg 1 $ and $k \in N$.
We find that the TRM gives the same transition frequency of $\w_c = 2 \pi  T_H ^s  =1/(4 M)$ as one from QNM. 
In most cases,  QNM was enough in finding the horizon area spectrum of a black hole. But for some cases it is not sufficient. In particular for the black holes with two horizons, we find that other damped oscillation modes are necessary to obtain the spectra of the both inner and outer horizon areas. In our previous works \cite{sk, sk2}  we have seen that  the spectra of the both inner and outer horizon areas are obtained from the two transition frequencies. The Kerr black hole we consider in this paper  also has that property.
Throughout this paper, the units with $c=G=1$ are used.

%
\textit{Area and entropy spectra}--- 
The perturbations of black hole spacetimes are represented by the radial Schr$\ddot {\rm o}$dinger-like wave equations of the form 
\begin{equation} \label{weq}
 { {\partial ^2 f(z)} \over { \partial z}^2} + \left(\w^2 -V_z(z) \right) f(z) =0  ,
\end{equation}
where $z=z(r)$ is a tortoise coordinate which has the behavior of 
$z \sim r $ as $r \rightarrow \infty$ and $z \rightarrow - \infty$ as $r \rightarrow r_+$ with  outer horizon radius $r_+$. The linearized and massless perturbation of the Kerr black hole is described by Teukolsky's equation \cite{teu} which can be written in the form of the equation  (\ref{weq}).
The wave equation for the Kerr black hole can be solved in the highly damped regime by using WKB approximation along specific contours in the complex $r$-plane. The contours are anti-Stokes lines which is defined as $\rm Re\left[i \w (z- z' ) \right]=0$ with $z'=z (r')$ of some reference point. The excitations of the three anti-Stokes lines corresponds to QNM, TTM, and TRM \cite{keshk2}. 
Therefore the highly damped quasinormal modes(QNM) of the Kerr black hole are given by \cite{keshk2, keshk}
\begin{eqnarray} \label{skerr1}
\w^{QNM} \simeq  \tilde \w^{QNM}  - i  2 \pi T_H ^s \left(k+{1 \over 2} \right), 
\end{eqnarray}
where $\tilde \w^{QNM}$ is a function of black hole parameters, $k$ is overtone number with $k \in N, k \gg1$, and $T_H^s=1/(8 \pi M)$. $T_H^s$ is the Hawking temperature of the Schwarzschild black hole which has the same mass of the Kerr black hole. 
From this,  the transition frequency in the semiclassical limit is obtained; 
\begin{equation} \label{trank1}
\w^{QNM} _c =   ( | \w_ {I} | )_{k} - ( | \w_ {I} | )_{k-1} \simeq 2 \pi T_H ^s = {1 \over {4 M}} .
\end{equation}
Using Eq.(\ref{myf}) with this frequency, we have a following quantization condition: 
$\I^{QNM}= 2 M^2 = {A_{tot} \over {8 \pi}}= n_q \hbar $,
where $n_q =1,2,3,..$. 
Therefore we find that the total horizon area is quantized and equally spaced as follows:
\begin{equation} \label{skerr3}
A_{tot} =8 \pi n_q \hbar .
\end{equation}
The reason why only the total horizon area is quantized is because the transition frequency (\ref{trank1}) has no information of the angular momentum of the Kerr black hole. 
Therefore only with this quantization condition, it is hard to find the spectrum of the outer horizon area.
However as we propose, TTM and TRM as well as QNM should be considered. Therefore we will obtain transition frequencies corresponding to  TTM and TRM, and use them into the formula (\ref{myf}) to obtain quantization conditions. 

The discrete frequencies of QNM, TTM, and TRM in the highly damped regime are given by \cite{keshk2}
\begin{equation} \label{skerr2}
\w^j (k)=\tilde \w^j + i 4 \pi T^j \left(n+ { \mu^j \over 4} \right),  ~(n \in Z, \vert n \vert \gg1)
\end{equation}
where $T^j$ are some functions of black hole parameters and $\mu^j$ are  Maslov indices. The index $j$ indicates QNM, TTM, and TRM according to the specific contours.
The TTM and TRM modes can be obtained from the relations between the parameters $T^j$ and $\tilde \w^j$ of three resonance modes obtained  in \cite{keshk2};
\begin{eqnarray}
{1 \over {2 T^{TTM}}} - {1 \over {2 T^{QNM}}} ={1 \over {2 T^{TRM}}} &=& {1 \over  T_H}  ,\\
{ {\tilde \w^{TTM}} \over {2 T^{TTM}}} - { {\tilde \w^{QNM}} \over {2 T^{QNM}}} = { {\tilde \w^{TRM}} \over {2 T^{TRM}}} &=& {{m \Omega} \over T_H} +i 2 \pi s ,
\end{eqnarray}
where $\Omega$ is the angular velocity at horizon, $T_H$ is the Hawking temperature of the Kerr black hole and $s $ is spin of the fields, i.e. gravitational ($s$= $-2$), electromagnetic ($s$= $-1$), and scalar ($s$= $0$) fields.
Therefore, using the above relations and $T^{QNM} \simeq  -{{T_H^s} / 2}$, 
we find that the TTM and TRM are given by 
\begin{eqnarray}
\w^{TTM}&=& \tilde \w^{TTM} - i 2 \pi \left( { {T_H T_H^s \over {T_H^s -T_H}}} \right) (k - s)  ,\\
 \w^{TRM} &=& m \Omega - i 2 \pi T_H(k-s),
\end{eqnarray}
where
\begin{equation}
\tilde \w^{TTM}=\left( { { m \Omega T_H ^s - \tilde \w^{QNM} T_H} \over {T_H^s-T_H}} \right) ,
 \end{equation}
$m$ is the azimuthal harmonic indices, and $k$ is overtone number which is $k \in  N$ and $k \gg 1$ for highly damped modes.
In the highly damped regime we find that the TRM modes is the same as the  quasinormal modes (\ref{kqn}) obtained  in the previous work of Ref.\cite{hodk}.
From the TTM and TRM, we find the following transition frequencies corresponding to each mode:
\begin{eqnarray}
\w^{TTM}_c &=& 2 \pi \left({ {T_H T_H^s \over {T_H^s -T_H}}}\right) = - 2 \pi T_{in} \nonumber\\ 
&=& \frac{\sqrt{M^4-J^2}}{2 M (M^2  - \sqrt{M^4-J^2} ) }, \\
\w^{TRM}_c &=& 2 \pi T_H = \frac{\sqrt{M^4-J^2}}{2 M  \left(M^2+\sqrt{M^4-J^2}\right) },
\end{eqnarray}
where $T_{in}= {\kappa_-} / (2 \pi)$ with the surface gravity $\kappa_-$ at inner horizon which is negative \cite{muin}.
With these  transition frequencies, we obtain the quantization conditions from the formula (\ref{myf}); 
\begin{eqnarray} \label{kerraction}
 \I^{TTM} &=& \sqrt{M^4-J^2}-M^2 = - n_t \hbar , \\
\label{kerraction2}
 \I^{TRM} &=& M^2+\sqrt{M^4-J^2} = n_r \hbar ,
\end{eqnarray}
where $n_t$ and $n_r$ are positive integers.
Therefore, we find that the inner horizon area and outer horizon area are quantized as follows: \begin{eqnarray}\label{skerr4}
A_{in}=  8 \pi n_t \hbar ~~,~~A_{out}= 8 \pi n_r \hbar ~.
\end{eqnarray}
The inner horizon area spectra can be also obtained from the spectra of the total horizon area (\ref{skerr3})  and outer horizon area (\ref{skerr4}), i.e. $A_{in}=A_{tot}-A_{out} = 8 \pi (n_q - n_r) \hbar $. 
By comparing this to Eq.(\ref{skerr4}), we find that the relation between the quantum numbers is given by  $n_t= n_q - n_r >0$.
Therefore we find that the spectra of the both inner and outer horizon areas are equally spaced as $ \triangle A_{out} =\triangle A_{in} =8 \pi \hbar$.
By Bekenstein-Hawking area law \cite{area}, the entropy spectrum is also equally spaced; $\triangle S =2 \pi$.
In the previous work for the area spectra of the BTZ black hole, it was found that the two transition frequencies from the two families of the quasinormal modes lead to the quantization conditions of total horizon area and the difference between two horizon areas \cite{sk}. In this paper,  we find the same property for the Kerr black hole. While the transition frequency (\ref{trank1}) from the quasinormal modes (\ref{skerr1}) is associated with the quantization of the total horizon area, the transition frequencies from TTM and TRM, which are proportional to the temperatures of the inner and outer horizons,  lead to the quantizations of the inner horizon area and the outer horizon area, respectively. 
The quantization of the inner horizon area can imply that there may be  physical dynamics inside the outer horizon.  For example, the Hawking radiation might happen at inner horizon \cite{make}. To clarify the physical meaning of the inner horizon spectrum, the further investigation on the physical dynamics inside outer horizon is needed.

We can also consider the quantization of other black holes in this manner.  The three types of resonance modes can be obtained from the zeros and poles of transmission ($T$) and reflection ($R$) amplitudes for waves traveling from spatial infinity to the black hole horizon.  From the  transmission and reflection amplitudes in Ref.\cite{neitz}, for example, we find that the  Reissner-Nordstr{$\rm {\ddot {o}}$}m black hole has three resonance modes  of QNM, TTM, and TRM. But, the QNM cannot be obtained algebraically from the poles of transmission $T$ and reflection $R$, while the TTM and TRM are obtained as purely imaginary ones. 
Therefore by using TTM and TRM the spectra of the both inner and outer horizon areas can be obtained.
The analogs of Eqs.(\ref{kerraction}) and(\ref{kerraction2}) are given by 
\begin{eqnarray} 
 \I^{TTM} &=& M \sqrt{M^2-Q^2}-M^2 = - n_t \hbar , \\
 \I^{TRM} &=& M^2+M \sqrt{M^2-Q^2} = n_r \hbar .
\end{eqnarray}
It turns out that the TTM and TRM are associated with the quantizations of the inner horizon area and outer horizon area, respectively. 
It is easily found that the area and entropy spectra are given by $\triangle A_{out/in}=8 \pi  \hbar$ and $\triangle S=2 \pi$. This result is consistent with the area spectrum of the Reissner-Nordstr{$\rm {\ddot {o}}$}m black hole in the small charge limit \cite{ortega}.
Therefore we conclude that  these four dimensional black holes such as Schwarzschild, Kerr and Reissner-Nordstr{$\rm {\ddot {o}}$}m black holes have the universal behavior of  equally spaced entropy spectra as $\triangle S=2 \pi$.

\textit{Conclusion}---
We calculated the area and entropy spectra of the Kerr black hole. We noticed that it has three types of highly damped resonance modes which are  quasinormal modes (QNM),  total transmission modes (TTM) and total reflection modes (TRM). 
We proposed that all of these modes should be considered to carry information about quantum black hole since all they are  characteristic modes of  black hole, and therefore  they should be used in quantizing the black hole. 
Then the transition frequencies between the highly excited neighboring levels of the Kerr black hole were obtained from all of these modes. Based on Bohr's correspondence principle, the quantum black hole with a transition frequency at large quantum number is considered as the classical periodic system with the oscillation frequency equal to the transition frequency in the semiclassical limit. The action variable $\I$  of the classical system of periodic motion is identified and quantized via the Bohr-Sommerfeld quantization in the semiclassical limit as the formula (\ref{myf}).
By applying this method for the Kerr black hole, we obtained that the spectra of the inner and outer horizon area are equally spaced as $\triangle A_{out/in}=8 \pi  \hbar$, which is consistent with Bekenstein's proposal \cite{be}. Therefore the entropy spectrum also has equal spacing of $\triangle S=2 \pi$, which means that the Kerr black hole also has the universal behavior of equally spaced entropy spectrum like other black holes in Refs.\cite{sk, sk2}. 
In the same way, we also found that the Reissner-Nordstr{$\rm {\ddot {o}}$}m black hole has the universal behavior of  equally spaced entropy spectra as $\triangle S=2 \pi$.
These results agree with the quantization of the entropy spectrum obtained in different methods of Refs.\cite{pad, ropo, med2}. 
Our results also give good examples for the claim in  \cite{sk2} that there is the universality that the entropy spectrum of a black hole is equally spaced. Therefore  we found that the universality holds regardless of the dimensions, the presence of the angular momentum or charge, and gravity theory. 
It is expected that the universal behavior of entropy spectrum would be useful for understanding and investigating a quantum nature of black holes as the first step toward quantum gravity.


\section*{Acknowledgments}
\noindent This research  was supported by the National Research Foundation of Korea(NRF) grant funded by the Korea government(MEST) (No.2009-0063068) and also supported by Basic Science Research Program through the National Research Foundation of Korea(NRF) funded by the Ministry of Education, Science and Technology(No.2010-0008109).


\end{document}